\documentclass{PoS}
\usepackage{graphicx}
\usepackage{amsmath}
\usepackage{xcolor}
\usepackage{caption}
\usepackage{subcaption}

\title{Rho meson decay width in SU(2) gauge theories with 2 fundamental flavours.}

\ShortTitle{Rho meson decay width in SU(2) gauge theories with 2 fundamental flavours.}

\author{\speaker{Tadeusz Janowski},$^a$ Vincent~Drach$^{bc}$ and Claudio~Pica$^a$\\
\llap{a} CP3-Origins, University of Southern Denmark,\\
Campusvej 55, DK-5230 Odense M, Denmark\\
\llap{b} Centre for Mathematical Sciences, Plymouth University, \\
Plymouth, PL4 8AA, United Kingdom,\\
\llap{c} CERN, Theoretical Physics Department,\\
Geneva, Switzerland\\  
E-mail: \email{janowski@cp3.sdu.dk}, \email{vincent.drach@plymouth.ac.uk}, \email{pica@cp3.sdu.dk}}

\abstract{SU(2) gauge theories with two quark flavours in the fundamental representation are among the most promising theories of composite dynamics describing the electroweak sector. Three out of five Goldstone bosons in these models become the longitudinal components of the W and Z bosons giving them mass. Like in QCD, we expect a spectrum of excitations which appear as resonances in vector boson scattering, in particular the vector resonance corresponding to the rho-meson in QCD.

In this talk I will present the preliminary results of the first calculation of the rho-meson decay width in this theory, which is analogous to rho to two pions decay calculation in QCD. The results presented were calculated in a moving frame with total momentum (0,0,1) on two ensembles. Future plans include using 3 moving frames on a larger set of ensembles to extract the resonance parameters more reliably and also take the chiral and continuum limits.}

\FullConference{34th annual International Symposium on Lattice Field Theory\\
         24-30 July 2016\\
         University of Southampton, UK}

\begin{document}
\newcommand{\bra}[1]{\langle #1 \mid}
\newcommand{\ket}[1]{\mid\! #1 \rangle}
\newcommand{\vev}[1]{\langle 0 \mid\! #1 \mid\! 0 \rangle}
\newcommand{\matel}[3]{\langle #1 \mid\! #2 \mid\! #3 \rangle}
\newcommand{\cset}[1]{\mid\! #1 \rangle\langle #1 \mid\!}
\newcommand{\innerprod}[2]{\langle #1 \mid \! #2 \rangle}
\newcommand{\x}[0]{\mathbf x}
\newcommand{\msbar}[0]{\overline{\mathrm{MS}}}

\section{The model}
One of the most promising candidate models which can explain the electroweak symmetry breaking in a natural way is the SU(2) gauge model with two fundamental flavours \cite{Cacciapaglia:2014uja}.
An important feature of the fundamental representation of SU(2) is its pseudo-reality. This means that a field in an anti-fundamental representation can be converted into a field in the fundamental representation by the application of $C = i\sigma^2_{c}$ colour matrix. As a consequence, fermion fields $q$ and $i\sigma^2_s q^*$ (where $\sigma^2_s$ is a matrix in spinor space) can be combined into a single flavour multiplet. In the two-flavour case, this multiplet can be written explicitly as:
  \begin{equation}
    Q = \left( \begin{array}{c}u_L\\d_L\\ -i\sigma^2_s C \bar u_R^T\\ -i\sigma^2_s C \bar d_R^T\end{array} \right)
  \end{equation}
  which is symmetric under SU(4) flavour group.

  This flavour symmetry is broken by the formation of the condensate $\Sigma_{ab} = \langle Q_a (-i\sigma^2_s) C Q_b^* \rangle$, where $a$ and $b$ are flavour indices and spin and colour indices are contracted. Under flavour group transformations, the condensate transforms as $\Sigma \to u \Sigma u^T$ where $u$ is a matrix in a subgroup of the $SU(4)$ flavour group, which leaves the condensate invariant. This induces symmetry breaking pattern $SU(4) \to Sp(4)$. Because $SU(4)$ is 15-dimensional and $Sp(4)$ is 10-dimensional, this gives us 15-10=5 pseudo Nambu-Goldstone Bosons (pNGBs), which will be called ``pions'' in the remainder if this article, in analogy with the QCD case.

  To relate our model with the Standard Model Higgs sector, we need to impose $SU(2)_L \times U(1)_Y$ gauge symmetry on our fields. Thus, in analogy with the Standard Model, we combine the left-handed $u_L$ and $d_L$ into an $SU(2)_L$ doublet with hypercharge $Y=0$, while the right-handed fields remain $SU(2)_L$ singlets with hypercharges +1/2 and -1/2 respectively. We can see that the condensate is going to transform under the electroweak symmetry transformations as $\Sigma \to g \Sigma g^T$, where $g$ is a block-diagonal matrix of the form $SU(2) \oplus e^{i \alpha(x)} \oplus e^{-i\alpha(x)}$.
  We can see by inspection that, depending on the choice of the condensate $\Sigma$, the electroweak symmetry may or may not be broken. These options along with their consequences for pNGBs are summarised in Table \ref{tab:1}. The additional fields can be interpreted as dark matter candidates \cite{Lewis:2011zb}.

	\begin{table}
	  \centering
	\begin{tabular}[3]{c|cc}
	  & $\Sigma_B$ & $\Sigma_H$\\
	  \hline
	  EW symmetry & unbroken & broken \\
	  model & composite Higgs & Technicolor \\
	  pNGBs & $W^\pm$, $Z$, $H$ + 1 extra & $W^\pm$, $Z$ + 2 extra \\
	  Higgs & pNGB & scalar resonance 
	\end{tabular}
	\caption{Comparison between vacuum alignments in SU(2) model with 2 fundamental flavours.}
	\label{tab:1}
      \end{table}

	Even more interestingly, the vacuum can also be a superposition of the symmetry breaking vacuum $\Sigma_H$ and symmetry-conserving one $\Sigma_B$:
	\begin{equation}
	  \Sigma_0 = \cos \theta \Sigma_B + \sin \theta \Sigma_H
	\end{equation}
In this scenario, the physical Higgs boson is the superposition of the pNGB and the scalar techni-$\sigma$ state.
	The spectrum of the model has been studied in \cite{Arthur:2016ozw,Arthur:2016dir}.

	In this study, we are interested in the scattering of pNGBs, which is equivalent to the scattering of heavy vector bosons at high energies (equivalence theorem). This allows us to give predictions about the resonance spectrum in these processes.
	In our model, the pNGBs are in the 5-dimensional representation of $Sp(4)$. Because $Sp(4)$ is locally isomorphic to $SO(5)$, we can think of the pNGBs as being in the fundamental representation of $SO(5)$. This immediately leads to three possible irreducible representations of two-pion state:
      \begin{enumerate}
	\item 14-dimensional symmetric traceless: $\pi^i\pi^j + \pi^j\pi^i - \frac{2}{5} \pi^k\pi^k \delta^{ij}$ - see \cite{Arthur:2014zda}
	\item 10-dimensional antisymmetric $\pi^i\pi^j - \pi^j\pi^i$
	\item 1-dimensional trace $\delta^{ij}\pi^k\pi^k$.
      \end{enumerate}
      In this proceeding we study the antisymmetric case. It is easy to see in the centre-of-mass frame that the two-pion wavefunction is odd under parity. We can rewrite it in the partial-wave basis $\ket{E,P,l,m}$, where $E$ is the total energy, $P$ is the total momentum and $l$ and $m$ are the integers corresponding to total angular momentum and the angular momentum in the z-direction. The parity transformation of such a state is $(-1)^l$, which implies that $l$ is odd. We can then conclude, using angular momentum conservation, that this representation can mix with a vector ($J=1$) state.
      Note that this discussion is analogous with $(\pi\pi)_{I=1}$ case in QCD, which exhibits a vector resonance $\rho$.

	We want to investigate the techni-$\rho$ (abbreviated henceforth as $\rho$) resonance parameters, namely the resonance mass $m_\rho$ and width $\Gamma_\rho$. Equivalently, the width can be expressed in terms of the effective coupling $g_{\rho\pi\pi}$ defined as
	\begin{equation}
	  \mathcal L_{eff} = g_{\rho\pi\pi} \epsilon_{ijk} \rho^{i\mu}\pi^j\partial_\mu\pi^k 
	\end{equation}
	In QCD $g_{\rho\pi\pi}\approx 6$. This need not be the case for this model.
	The relation between $\Gamma_\rho$ and $g_{\rho\pi\pi}$ is:
	\begin{equation}
	  \Gamma_\rho = \frac{g_{\rho\pi\pi}^2}{6\pi}\frac{p^3}{m_\rho^2}\qquad p = \sqrt{m_\rho^2/4 - m_\pi^2}.
	\end{equation}

  Below inelastic threshold ($4m_\pi$) there is only one state contributing to the S-matrix, which can be written as:
  \begin{equation}
    S = \innerprod{E,p,l',m'}{E_{CM},0,l,m} = \delta(E-E_{CM}) \delta(p) \delta_{ll'} \delta_{mm'} e^{2 i \delta_l(E_{CM})}
  \end{equation}
  This defines the phase shift $\delta_l(E_{CM})$,which can then be related to $g_{\rho\pi\pi}$ using
  \begin{equation}
    \tan \delta_1 = \frac{g_{\rho\pi\pi}^2}{6\pi}\frac{p^3}{E_{CM}(m_\rho^2-E_{CM}^2)}\qquad p = \sqrt{m_\rho^2/4 - m_\pi^2}.
  \end{equation}

\section{Scattering on a lattice}

As discussed in the previous section, our calculation is analogous to $\rho\to\pi\pi$ scattering in QCD, which has been extensively studied on the lattice \cite{rpp1,rpp2,rpp3,rpp4,rpp5,rpp6,rpp7,rpp8,rpp9}.
  Phase shifts can be calculated from the energy spectrum using the approach first described by L\"uscher \cite{Luscher:1990ux} and generalised to moving frames by Rummukainen and Gottlieb \cite{Rummukainen:1995vs}.
  The idea is to exploit finite volume effects to relate the energy spectrum of two-pion states to the phase shift.
  The exact formula depends both on the reference frame and the representation of the interpolating operators in that frame.

  Specifically, we are interested in $l=1$ states in the infinite volume, which have a wavefunction proportional to the spherical harmonic $Y_{1m}(\theta,\varphi)$. However, in the finite volume we no longer have a full rotational symmetry. If we are in the centre-of-mass (COM) frame, then the symmetry is broken to the cubic symmetry group $O$. 
  The $l=1$ spherical harmonic then corresponds to the irreducible representation $T_1^-$, which unsurprisingly is the representation of a vector.

  Moving frames offer additional complication. This is because the phase shift is defined in the COM frame, which means that our system needs to be boosted back to the COM frame.
  Then, because of Lorentz contraction, the symmetry group will be reduced.
  We consider two moving frames: MF1 with total momentum (0,0,p) and MF2 with total momentum (p,p,0).
  In MF1 the cube is contracted along one of its sides resulting in a cuboid. The symmetry group describing this object is the dihedral group $D_{4h}$.
  The $Y_{1m}(\theta,\varphi)$ transforms reducibly under this group and Clebsch-Gordan decomposition reveals that it can be written as a direct sum of 1-dimensional $A_2^-$ representation and 2-dimensional $E^{-}$ representation. Either of the two can be used to extract the $l=1$ partial wave. In this work we use the $A_2^-$ representation.
  Finally the MF2 respects the $D_{2h}$ symmetry, under which $Y_{1m}(\theta,\varphi)$ reduces to three 1-dimensional representations $B_1^- \oplus B_2^- \oplus B_3^-$. 

  The formulae for the $l=1$ phase shift are given in Table \ref{tab:ps}. They are given as functions of the L\"uscher zeta function given by:
  \begin{table}
    \centering
  \begin{tabular}{c|c|c}
    frame & representation & $\tan \delta_1$\\
    \hline
     COM & $T_1^-$ & $\frac{\pi^{3/2}q}{Z_{00}(1;q^2)}$\\
      MF1 & $A_2^-$ & $\frac{\pi^{3/2}q}{Z_{00}(1;q^2) + \frac{2}{\sqrt 5 q^2} Z_{20}}$\\
      MF2 & $B_1^-$& $	\frac{\pi^{3/2}q}{Z_{00}(1;q^2) - \frac{1}{\sqrt 5 q^2} Z_{20} +i\frac{\sqrt{3}}{\sqrt{10}q^2}\left( Z_{22}(1;q^2) - Z_{2(-2)}(1;q^2) \right)}$
  \end{tabular}
  \caption{Expressions for $\delta_1$ phase shift calculated in different moving frames.}
  \label{tab:ps}
\end{table}
  \begin{equation}
    Z_{lm}(s,q^2) = \sum_{n\in \mathbb{Z}^3}\frac{Y_{lm}(n)}{(q^2-n^2)^s}
  \end{equation}
  where $q^2$ depends on the centre-of-mass energy of the pion system via the continuum dispersion relation formula
  \begin{equation}
    E_{CM} = 2 \sqrt{m_\pi ^ 2 + \left(\frac{2 \pi q}{L}\right)^2}.
    \label{eq:qdef}
  \end{equation}
  
  The problem is then reduced to accurately obtaining the energy levels of the two-pion system.
  This can be done by analysing the time dependence of the two-point correlation function, which is (assuming total time extent $T>>t$):
  \begin{equation}
    C_{ij}(t) \equiv \matel{0}{O_i^\dagger(t) O_j(0)}{0} = \sum_{n,m} \matel{0}{O_i^\dagger}{n} (e^{-E_nt}\delta_{mn}) \matel{m}{O_j}{0},
    \label{eq:corrfn}
  \end{equation}
  where the operators $O_{i/j}$ have the same quantum numbers as the state we are trying to create, i.e. same representation of the cubic/dihedral group and 10-dimensional representation of the flavour symmetry group.
We then see that for large time separations the exponential suppresion will remove all but the lowest energy state.
In practice this procedure is not good enough for studying resonances. This is because in order to have sensitivity to the resonance we require the two-pion energy to be close to the resonance energy. This means that the lowest two energy states in the spectrum will be very close together and extracting the ground state would require going to prohibitively large values of $t$.

The solution is to use the generalised eigenvalue problem (GEVP) approach. In this approach we use as many interpolating operators as the number of energy levels we want to extract and define square matrices in Eq. \ref{eq:corrfn} $U_{in}\equiv \matel{0}{O_i^\dagger}{n}$ and $V_{mj} \equiv\matel{m}{O_j}{0} $.
  Then, assuming that $t$ is large enough that higher energy levels don't contribute, we have:
  \begin{equation}
    C_{ij}^{-1}(t_0) C_{jk}(t) = V^{-1}_{in} diag \left( e^{-E_n(t-t_0)} \right)_{nm} V_{mj}.
  \end{equation}
  The spectrum can then be extracted from the eigenvalues of $C^{-1}(t_0)C(t)$.

  We use the following two interpolating operators
  \begin{align}
    O_1(t) &= \sum_{x,y} \bar \psi(x) \gamma^5 \psi(x) \bar \psi(y)\gamma^5 \psi(y) e^{i \mathbf p \cdot \mathbf x}\\
    O_2(t) &= \sum_x \bar \psi(x) (\gamma \cdot \hat p)  \psi(x) e^{i \mathbf p \cdot \mathbf x}
  \end{align}

  The corresponding contractions are:
	\begin{align}
	  C_{11}(t) &= \vcenter{\hbox{\includegraphics[width=0.2\textwidth]{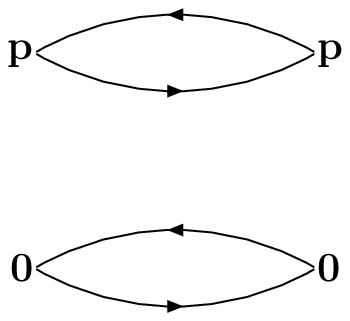}}}-\vcenter{\hbox{\includegraphics[width=0.2\textwidth]{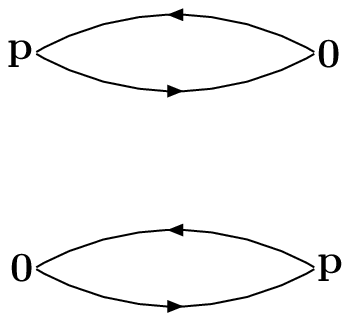}}}+\vcenter{\hbox{\includegraphics[width=0.2\textwidth]{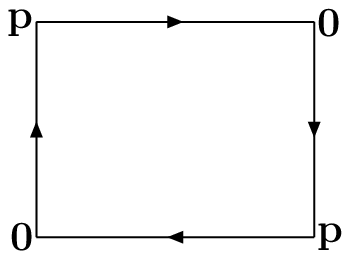}}}\nonumber\\
	  &+\vcenter{\hbox{\includegraphics[width=0.2\textwidth]{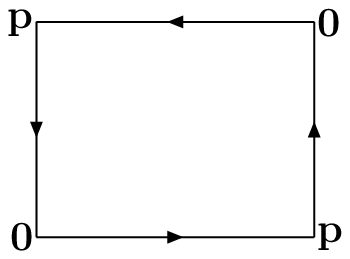}}} -\vcenter{\hbox{ \includegraphics[width=0.2\textwidth]{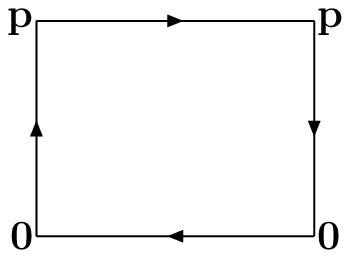}}} 
	  - \vcenter{\hbox{\includegraphics[width=0.2\textwidth]{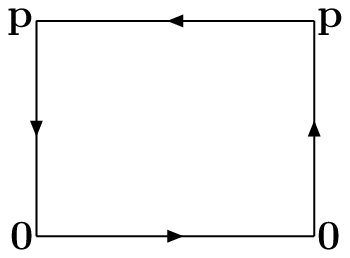}  }}
	\end{align}
	\begin{align}
	  C_{12}(t) &=-C_{21}^*(t)=\vcenter{\hbox{\includegraphics[width=0.2\textwidth]{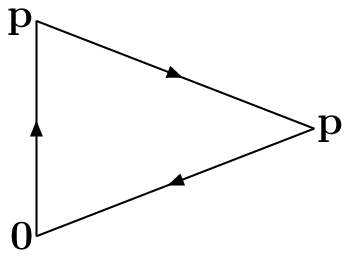}  }} - \vcenter{\hbox{\includegraphics[width=0.2\textwidth]{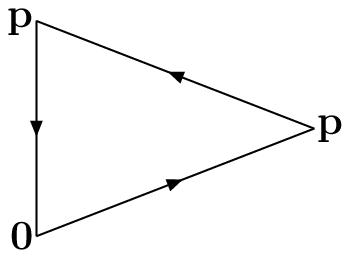}  }}
	\end{align}
	\begin{align}
	  C_{22}(t) &=\vcenter{\hbox{\includegraphics[width=0.2\textwidth]{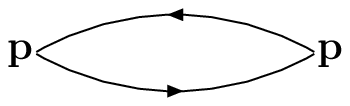}  }}
	\end{align}

	Each of these diagrams can be calculated on a lattice. For the ``direct'' diagrams, where two pions propagate without exchanging any quarks, we use propagators calculated at time slice $0$. Triangle and rectangle diagrams require one inversion per timeslice, making the calculation particularly expensive. We use sequential propagators to ensure the $\sum_x \bar \psi(x)\psi(x)$ structure at the intermatiate vertices.
	Finally, to ensure this structure at the source timeslice we use the stochastic noise ``one-end trick'', where we rewrite the operator at the source as $\sum_{x,y}\langle \bar \psi(x) \eta^\dagger(x) \eta(y)\psi(y) \rangle_\eta$, where we introduced the stochastic random noise $\eta(x)$ satisfying $\langle \eta^\dagger(x)\eta(y)\rangle_\eta = \delta(x-y)$. We then invert the fermion propagators using $\eta(x)$ as the source and average over a number of $\eta$ configurations (called ``hits''). Note that in order to prevent unwanted contractions, we need a different noise source for each of the pions at the source.

\section{Results}
\begin{figure}
    \centering
    \begin{subfigure}{0.5\textwidth}
      \centering
      \includegraphics[width=\linewidth]{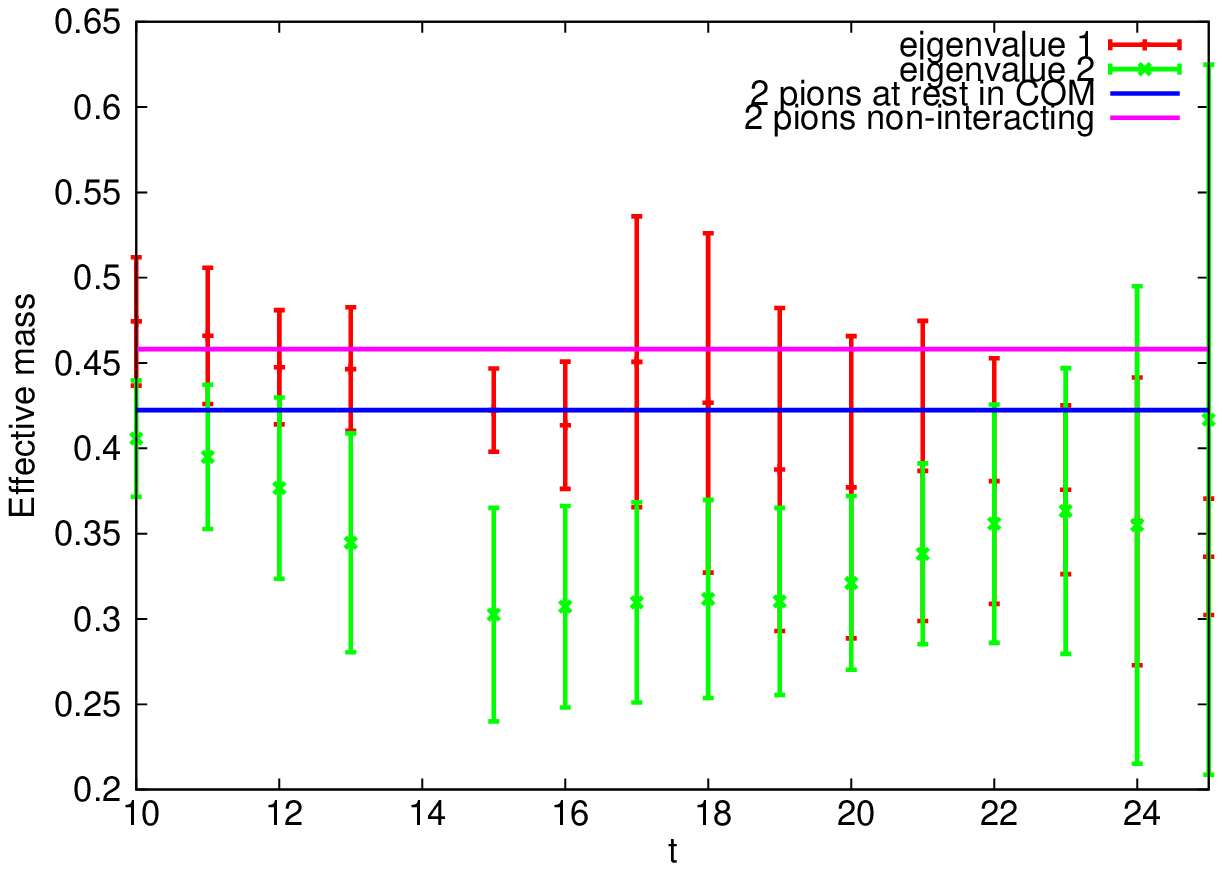}
    \end{subfigure}%
    \begin{subfigure}{0.5\textwidth}
      \centering
      \includegraphics[width=\linewidth]{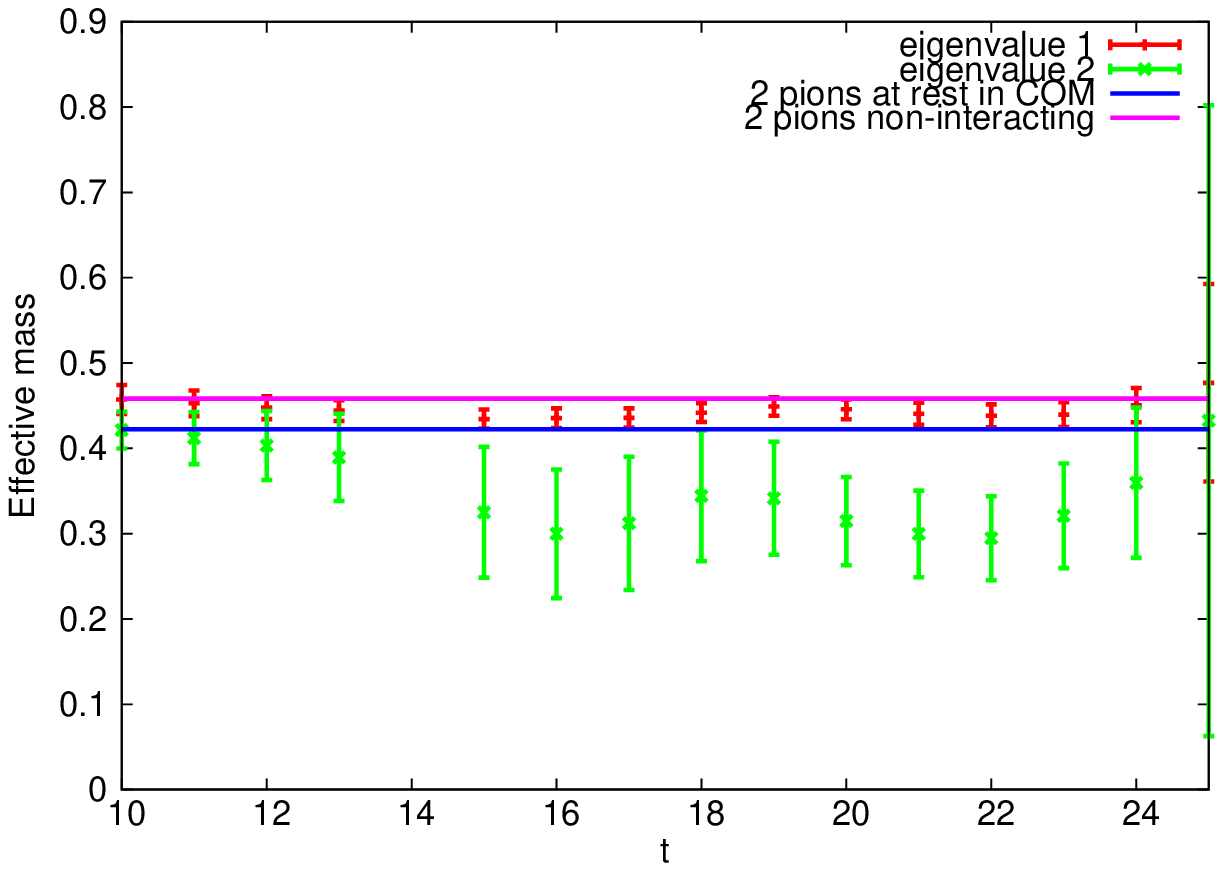}
    \end{subfigure}
    \caption{``Effective mass'' plots corresponding to the eigenvalues of the correlation matrix on two ensembles - $32^4$ with P+A boundary conditions (left) and $32^3\times 64$ with periodic boundary conditions (right).}
    \label{fig:effm}
  \end{figure}

The results presented in this section were obtained using two Wilson fermion ensembles with dimensions $32^4$ (122 configurations) and $32^3\times 64$ (112 configurations).
Both ensembles have the same parameters: $\beta = 2.0$, bare quark mass $a m_0 = -0.958$, which correspond to $a f_\pi = 0.049(3)$, $a m_\pi = 0.18(1)$ and $a m_\rho = 0.38(5)$. On the $32^4$ ensemble we use periodic+antiperiodic boundary conditions for the fermion fields to effectively double the time extent. The plots in Fig. \ref{fig:effm} show ``effective masses'' corresponding to each of the eigenvalues, defined as $E_i(t) = -\ln \lambda_i(t)/(t_0 - t)$. The upper horizontal line on both plots corresponds to the energy of two non-interacting pions. We expect the energy of the interacting pion pair to be lower, because the interaction in this channel is attractive. The lower line corresponds to the rest energy of two pions. 
Below this line $q$ defined in Eq. \ref{eq:qdef} becomes imaginary and, as discussed in \cite{Luscher:1990ux}, the quantity obtained with our procedure would not correspond to the phase shift.
If the $\rho$ state is unstable we expect the lower eigenvalue to be between the two horizontal lines. This is not clear from the plot.

  \section{Conclusions}
  We have presented the first attempt at calculating $\rho$ resonance mass and decay width in a non-QCD theory.
  The early results show no indication of $\rho$ meson being stable on the ensembles used and hence no information about the phase shift could be extracted.
  Future work in this area involves generating configuration where $\rho$ is unstable and adding the $O(a)$ improvement to the fermion action in hopes of getting a signal. In the long run we also plan on including the centre-of-mass frame in the analysis as well as performing the continuum and chiral extrapolations.

  {\bf Acknowledgements} This work was supported by the Danish National Re- search Foundation DNRF:90 grant and by a Lundbeck Foundation Fellowship grant. The computing facilities were provided by the Danish Centre for Scientific Computing and the DeIC national HPC center at SDU.

\end{document}